\documentclass[aps,prl,twocolumn]{revtex4-1}
\usepackage[bookmarks,colorlinks,citecolor=blue,linkcolor=blue]{hyperref}

\usepackage{amssymb}
\usepackage{amsmath,bm}
\usepackage{graphicx,color}
\usepackage{bbm}
\usepackage[bottom]{footmisc}
\usepackage{multirow}

\begin{document}
\title{Tuning the Viscosity and Jamming Point in Dense Active non-Brownian Suspensions}
\author{Bhanu Prasad Bhowmik}
\email{bhowmikbhanuprasad592@gmail.com}
\affiliation{School of Engineering, University of Edinburgh, Edinburgh EH9 3JL, United Kingdom}

\begin{abstract} 

Using numerical simulations, we study the rheological response of dense non-Brownian suspensions containing active particles. The active particles are modelled as run-and-tumble particles with three controlling parameters: the fraction of active particles in the system, an active force applied to the particles, and a persistence time after which the direction of the active force changes randomly. Our simulations reveal that the presence of activity can reduce the viscosity (by an order of magnitude) by decreasing the number of frictional contacts, which also shifts the jamming point to a higher volume fraction. Moreover, the microscopic structure of force chains in the presence of activity is qualitatively different from that in the passive system, showing reduced anisotropy. We also find that while the presence of activity drives the system away from jamming by preventing the formation of force chains, unjamming an already jammed state by breaking existing force chains requires a higher activity strength. Finally, we propose a new constitutive law to describe the rheology of dense active non-Brownian suspensions.


\end{abstract}

\maketitle

\paragraph{Introduction.}
Dense suspensions such as slurries, cement, blood, and molten chocolate are ubiquitous in daily life as well as in industry~\cite{Guazzelli_Pouliquen_2018, StickelAnnualReview, NessReview}. These materials are composed of two components, a suspending viscous fluid with viscosity $\eta_f$ and solid particles that have the same density $\rho$ as the fluid. The viscosity of such systems is controlled by various properties of the solid particles and exhibits a complex nonlinear growth with increasing solid volume fraction $\phi$, eventually leading to a divergence of viscosity at a property-dependent critical volume fraction $\phi_J$. This increase in viscosity, however, poses a challenge in various industrial processes, as the flow of the suspension becomes increasingly sluggish and energy-consuming.

Recently, researchers have focused on the problem of tuning the viscosity of dense suspensions using various methods. One common approach involves altering the properties of the solid particles by varying their surface smoothness, asphericity, or by adding lubricating chemicals to the system~\cite{HRZS_Isa_2018_PNAS, FernandezPRL2013, PRR_TunableThick, James2019}. However, these methods are not practical for implementation during the flow of a pre-prepared suspension. Another approach involves tuning the viscosity by applying external mechanical perturbations~\cite{superPosNatmat2010, Blanc_Lemaire_Peters_2014, LNCSC_PNAS_2016} during flow in a fixed direction, or by appropriately changing the flow direction itself~\cite{MariShearRotationPRL2023, Pappu_PRR_2024, Pappu_ArXiv2025}. Microscopically, such external perturbations reduce the number of direct frictional contacts between particles, whose proliferation with increasing $\phi$ is the origin of the rapid increase in viscosity. Recent studies suggest that cyclic deformation applied orthogonally to the primary flow can drive the system toward an absorbing state transition characterised by a vanishing number of particle contacts~\cite{NMC_ScAdv_2018}, whereas more general orthogonal perturbations may not necessarily lead to a transition but can still break existing contacts and prevent the formation of fragile force chains~\cite{CatesPRL1998}.

In the context of viscosity reduction, an interesting class of systems is dense suspensions composed of active particles~\cite{SriramReview,RevModPhys.88.045006}. By active particles, we refer to particles that can generate motion by dissipating their internal energy. Such systems are prevalent in biology, with a common example being various microswimmers suspended in a viscous fluid, such as bacterial suspensions~\cite{SokolovPRL2009}. Recently, artificially prepared active particles~\cite{Bocquet_PRL_2010, ActiveColloids_PRL_2007, JanusParticle_PRL_2016, Pannacci2007, Walther2013, QuinckeBasic1984, QuinckePRE2014, QuinckeOscillationPRL2021} have emerged as model systems to study their properties. These systems are primarily composed of active colloids, where the dynamics of solid active particles such as Janus or Quincke particles can be controlled by applying an external force. The rheology of microswimmer suspensions has been extensively studied at low volume fractions, showing a significant reduction in viscosity in the presence of active particles~\cite{SokolovPRL2009, clement2016bacterial, EricClement_PRL_2013, MSYD_PRL_2016,Choudhary2023}.From the perspective of numerical studies, most have been limited to Brownian soft particles or far from $\phi_J$~\cite{HurtmutSoftMatter,WieseKroyLevisPRL2023, ghosh2025influenceactivebreathingrheology}. However, how activity-induced dynamics affect the viscosity of non-Brownian suspensions near $\phi_J$ remains poorly understood. Moreover, recent studies suggest that the presence of active particles and the application of external cyclic deformation have similar effects on the rheology of the system, where active dynamics can act as a form of local driving~\cite{RishabSmarajitNatPhy2025, Goswami2025}. In the same spirit, one should expect a similar reduction in viscosity observed in Ref.~\cite{LNCSC_PNAS_2016} due to external perturbation in a similar context of micro-rheology~\cite{MicroRheo_1,MicroRheo_2,MicroRheo_3}. 

In this paper, using DEM simulations, we study the rheology of active dense suspensions of non-Brownian, frictional particles with different particle friction coefficients $\mu_p$, under an overdamped, rate-independent regime~\cite{BoyerGuazelliPouliquenPRL2011}. The active particles are modelled as non-Brownian run-and-tumble particles~\cite{RituparnoSoftMatter}, characterised by three parameters: the fraction of active particles in the system $C_a$, the active force exerted on the active particles $F_a$, and the persistence time $\tau_p$, which represents the duration for which the active force is applied in a constant direction. We find that the relative viscosity $\eta_r = \sigma_{xy}/\dot{\gamma}\eta_f$ decreases monotonically with $C_a$, but exhibits a non-monotonic dependence on $F_a$ and $\tau_p$. As $F_a$ and $\tau_p$ increase, $\eta_r$ initially decreases to a $\phi$-dependent minimum viscosity, $\eta_r^m(\phi)$, at specific values of active force $F_a^m$ and persistence time $\tau_p$, and then begins to increase. Here, $\sigma_{xy}$ and $\dot{\gamma}$ denote the shear stress and shear rate, respectively. We demonstrate that both $F_a^m$ and $\eta_r^m$ diverge as the system approaches the isotropic jamming volume fraction $\phi_{\text{RCP}}$~\cite{Phi_rcpFirst}. The presence of active particles alters the microscopic nature of the force chain network. The $\mu_p$ dependent shear jamming volume fraction, $\phi_J^{\mu_p}$, can be significantly tuned by varying any one of these parameters. We also find that the rheology of dense active suspensions resembles that of inhomogeneously driven systems. Finally, we establish a new constitutive law that holds for both passive and active dense suspensions. 
\paragraph{Simulation details.}
We simulate a bidisperse system of $N = 10^3$ granular particles with radii $R = a$ and $1.4a$ in a 50:50 number ratio, contained in a cubical simulation box of length $L$, with periodic boundary conditions applied~\cite{LeesEdwards_PBC__1972}. The particles are suspended in a density-matched ($\rho = 1$) fluid of viscosity $\eta_f = 0.1$ and experience three types of interactions. A particle with velocity $\bm{u}_i$ and radius $a'$ experiences a drag force $\bm{F}^d_i = 6\pi\eta_f a' (\bm{U}^{\infty} - \bm{u}_i)$, where $\bm{U}^{\infty}$ is the velocity of the background fluid at the particle’s position. The motion of one particle influences others via the viscous fluid by exerting hydrodynamic forces and torques, which depend on the interparticle distance $h_{i,j}$ and the relative velocity $\bm{u}_{i,j}$~\cite{KimAndKarila,BALL1997444}. The leading-order terms of these interactions are given by $\bm{F}^{h}_{i,j} \sim \frac{1}{h_{i,j}} (\bm{u}_{i,j} \cdot \bm{n}_{i,j})\bm{n}_{i,j}$ and $\bm{\tau}^{h}_{i,j} \sim \ln\left(\frac{a'}{h_{i,j}}\right)(\bm{u}_{i,j} \times \bm{n}_{i,j})$, where $\bm{n}_{i,j}$ is the unit vector pointing from the $j^{\text{th}}$ to the $i^{\text{th}}$ particle, and $a'$ is the radius of the smaller particle between the $j^{\text{th}}$ and $i^{\text{th}}$ particles. For simplicity, the lubrication interaction is set to zero when $h_{i,j} > a'/2$, and is regularised by setting $h_{i,j} = 10^{-3} \times a'$ when $h_{i,j} < 10^{-3} \times a'$, to avoid divergence and allow particles to come into contact~\cite{ChealAndNess}. The contact force and torque are modelled as $\bm{F}_{i,j}^{c} = k_n \delta_{i,j} \bm{n}_{i,j} - k_t \bm{\xi}_{i,j}$ and $\bm{\tau}^{c}_{i,j} = a_i (\bm{n}_{i,j} \times k_t \bm{\xi}_{i,j})$, where $k_n$ and $k_t$ are the normal and tangential stiffnesses, respectively, both set to $10^4$ to allow for a sufficiently large time step while keeping particle overlaps minimal. Here, $\delta_{i,j}$ is the overlap between the particles and $\bm{\xi}_{i,j}$ is the accumulated tangential displacement between particles, computed from the moment they come into contact until the contact is broken. This displacement accounts for the history dependence of the frictional force~\cite{CundallAndStrackDEM}. According to Coulomb’s criterion, the maximum allowable tangential force for frictional particles is given by $k_t \xi_{i,j} < \mu_p k_n h_{i,j}$. We simulate two different systems with $\mu_p = 0$ and $1$. To deform the system, we impose a velocity field on the background fluid given by $\bm{U}^{\infty}(y) = \dot{\gamma} y \hat{\bm{x}}$, where $y$ is the coordinate of the particle in the velocity gradient direction and $\hat{\bm{x}}$ is the unit vector along the flow direction. To ensure the system remains in the rate-independent regime, we choose a strain rate $\dot{\gamma} = 10^{-3}$ such that $\dot{\gamma} \ll \eta_f/\rho a^2$, $\dot{\gamma} \ll \sqrt{k_n/\rho a^3}$ and $\dot{\gamma} \le \tau_p^{-1}$~\cite{BoyerGuazelliPouliquenPRL2011, RomainMariPNAS2015}. Although, in our simulation setup, the natural unit of length, force and time are $a$, $k_n a$ and $\dot{\gamma}$, for clarity and consistency with simulation inputs, all the values reported in this paper are presented in their raw (dimensional) form and have not been rescaled by the unit. 
We randomly select $NC_a$ particles from the unsheared state and designate them as active particles. In addition to other forces, each active particle experiences an additional force of constant magnitude $F_a$, with direction changing independently after a time interval $\tau_p$. This induces self-propulsion in the viscous fluid in the absence of interactions. The shear stress $\sigma_{sxy}$ comprises three contributions: contact force, lubrication force, and drag force. Additionally, in active systems, there is an active stress given by $\sigma_{\alpha \beta}^a = -\frac{1}{V} \sum_{i = 0}^N f_{a,i}^{\alpha} r^{\beta}_i$, which vanishes upon ensemble averaging. This occurs because $\tau_p$ is chosen to be sufficiently small to prevent any directional ordering in the system~\cite{RituparnoPNAS2021}.       
\begin{figure*}
\includegraphics[width=0.97\textwidth]{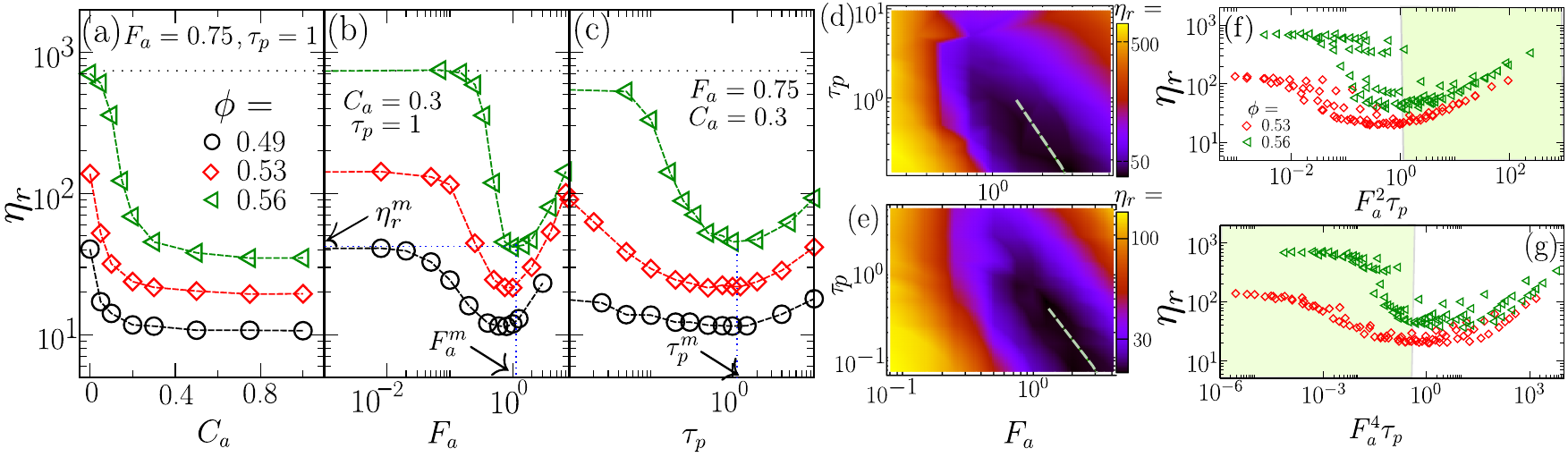}
\caption{
Tuning the viscosity by varying different parameters of active particles. Shown is the variation of relative viscosity $\eta_r$ as a function of (a) active particle fraction $C_a$, (b) active force $F_a$, and (c) persistence time $\tau_p$ for three different volume fractions $\phi$ below the jamming volume fraction $\phi_J^{\mu_p}$, for a particle friction coefficient $\mu_p = 1$. The black dotted line represents the viscosity of the passive system with $\phi = 0.56$. Green dotted lines show the coordinate of the minimum of $\eta_r$ - $F_a$ or $\eta_r$ - $\tau_p$ curve. Variation of $\eta_r$ with $F_a$ and $\tau_p$ at fixed $C_a = 0.3$ is shown for volume fractions $\phi = 0.56$ (d) and $0.53$ (e). Green dotted lines indicate the locus of the lowest viscosity points for fixed $F_a$ but different $\tau_p$. (f)--(g) $\eta_r$ expressed as a function of a combined parameter involving $F_a$ and $\tau_p$.
}
\label{fig1}
\end{figure*}
\begin{figure}
\includegraphics[width=0.47\textwidth]{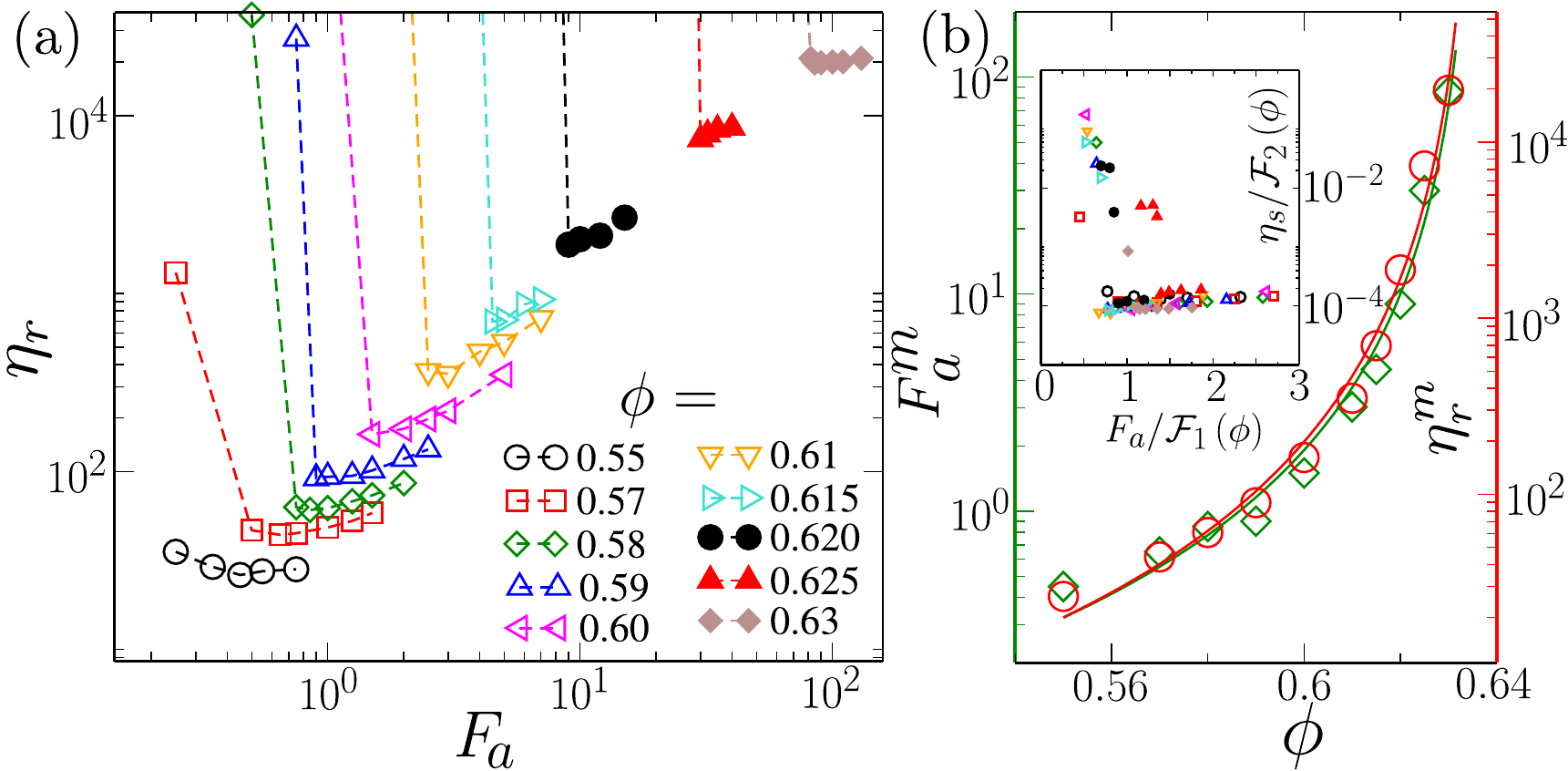}
\caption{(a) Relative viscosity $\eta_r$ as a function of active force $F_a$ for different volume fractions $\phi$ above and below $\phi_J^{\mu_p = 1}$. (b) The divergence of active force $F_a^m$ that yields minimum viscosity $\eta_r^m$, and minimum viscosity $\eta_r^m$ as $\phi \rightarrow \phi_{\text{RCP}}$. Green and red solid lines stand for the scaling functions $\mathcal{F}_1(\phi)$ and $\mathcal{F}_2(\phi)$, respectively. Inset: Collapse of data shown in (a) with the help of $\mathcal{F}_1(\phi)$ and $\mathcal{F}_2(\phi)$.}
\label{fig2}
\end{figure}
\paragraph{Tuning the viscosity using active particles.}
First, we focus on how $\eta_r$ depends on different parameters of active particles by running simulations for varying values of one parameter while keeping the other two parameters fixed, for three different $\phi$ values (Fig.~\ref{fig1}). We find that $\eta_r$ decreases monotonically with $C_a$ for all $\phi$ values, but exhibits a non-monotonic dependence on $F_a$ and $\tau_p$ (shown in Fig.~\ref{fig1}(a), (b) and (c)). In the case of dependence on $F_a$, $\eta_r$ remains more or less constant at small $F_a$, then decreases with $F_a$ to a $\phi$ dependent minimum viscosity $\eta_r^m$ (shown by blue dotted line) at a $\phi$ dependent active force $F_a^m$. This suggests that when the active force is too small, it's not able to break the force chains or prevent their formation and does not contribute significantly to particle diffusion. In this regime system behaves like a passive system. Only when $F_a$ exceeds a critical value does the system behave like an active system. When $\tau_p$ is varied similar response is observed with minimum viscosity at $\tau_p^m$, except with no clear plateau at small $\tau_p$. When $F_a > F_a^m$ and $\tau_p > \tau_p^m$, the trend reverses, with $\eta_r$ increasing with both $F_a$ and $\tau_p$. The viscosity reduction is more than an order of magnitude for the highest $\phi$. In Fig.~\ref{fig1}(d) and (e), the colour map of $\eta_r$ is shown as a function of $F_a$ and $\tau_p$ for two different values of $\phi = 0.56$ and 0.53, respectively, with a fixed $C_a = 0.3$. Here $F_a$ is chosen larger than the critical value below which no variation of $\eta_r$ is observed as a function of active force. It is clear that when $F_a$ is small, $\eta_r$ is weakly dependent on $\tau_p$ within the studied range. However, as $F_a$ increases, the effect of $\tau_p$ becomes prominent, with $F_a^m$ showing strong dependence on $\tau_p$. This observation suggests that although both $F_a$ and $\tau_p$ control $\eta_r$, however, they do not have equal weight. 

To understand with what weight $F_a$ and $\tau_p$ contribute to $\eta_r$, we introduce an effective temperature $T_{\text{eff}}$ in our non-Brownian active dense suspensions, originating from the motion of particles due to active forcing. An effective temperature for Brownian active systems of run-and-tumble particle is usually defined as $T_{\text{eff}} \sim F_a^2 \tau_p$~\cite{Bocquet_PRL_2010, Marchetti2014SoftMatter, RituparnoSoftMatter, KallolPNAS2023}. 
If this definition holds in our case, we should observe the same value of $\eta_r$ for simulations with different combinations of $F_a$ and $\tau_p$ that yield the same $T_{\text{eff}}$. In (f), $\eta_r$ vs. $F_a^2 \tau_p$ is shown for a fixed value of the latter using multiple combinations of $\tau_p$ and $F_a$. Although this definition of $T_{\text{eff}}$ appears to work at large $T_{\text{eff}}$, it fails to collapse data at smaller values. This is expected since at small $F_a$, $\eta_r$ is weakly dependent on $\tau_p$ within the studied range. Therefore, in the small $F_a$ regime, diffusion is strongly dominated by $F_a$ compared to $\tau_p$. To account for this, we introduce a new definition of the effective temperature: $T_{\text{eff}} \sim F_a^4 \tau_p$. The dependence of $\eta_r$ on $F_a^4 \tau_p$ is shown in (f), which successfully collapses the data in the small-value regime. However, the microscopic origin of this form of effective temperature with an exponent 4 is not studied here. This suggests that the rheology of active non-Brownian particles cannot be directly compared with that of Brownian passive particles by defining an effective temperature based solely on activity.  

\begin{figure}
\includegraphics[width=0.45\textwidth]{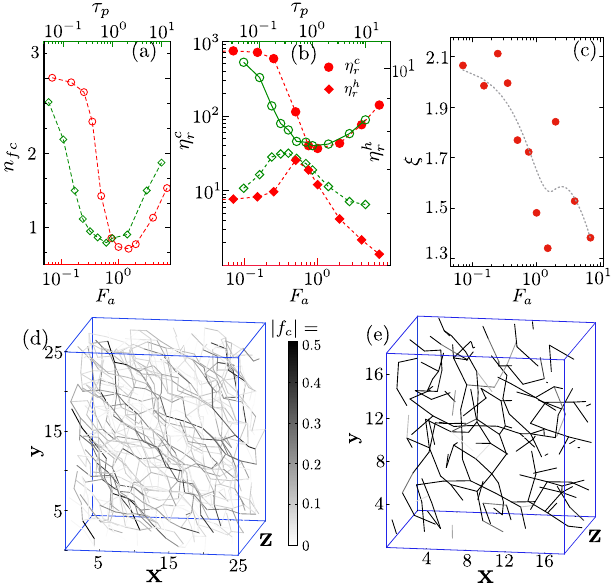}
\caption{Microscopic mechanism underlying the reduction of viscosity due to activity.
(a) Variation of the mean number of frictional contacts, $n_{fc}$, with active force $F_a$ (in red) and persistence time $\tau_p$ (in green).
(b) Variation of the viscosity contributions from frictional contacts, $\eta_r^c$ (solid symbols), and hydrodynamic interactions, $\eta_r^h$ (open symbols), with $F_a$ and $\tau_p$.
(c) Variation of the velocity–velocity correlation length, $\xi$, as a function of $F_a$. The black dotted line is the interpolation using red points for the eye guide.
(d)–(e) Force chain networks for a passive system with $\phi = 0.49$ and an active system with $\phi = 0.56$, $C_a = 0.3$, $F_a = 0.75$ and $\tau_p = 1$ which have  same $\eta_r$ as passive system.}
\label{fig3}
\end{figure}

\begin{figure}
\includegraphics[width=0.45\textwidth]{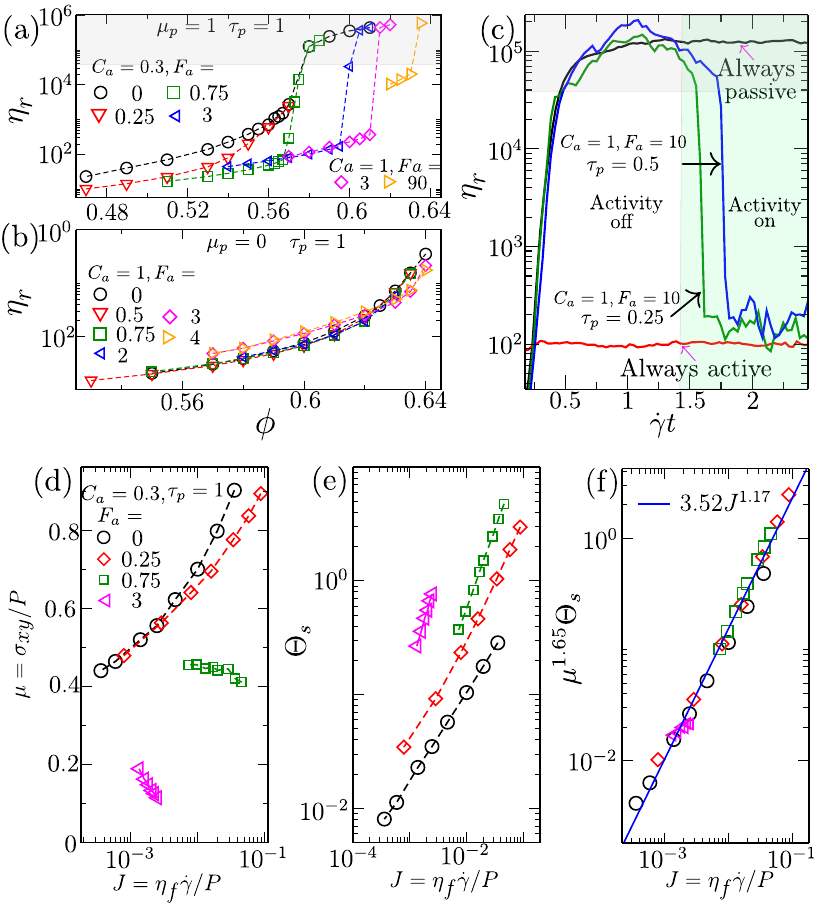}
\caption{Tuning the shear jamming volume fraction using active particles.
(a)-(b)  Relative viscosity, $\eta_r$, plotted as a function of volume fraction $\phi$ for various activity strengths, for $\mu_p = 1$ and $\mu_p = 0$.
(b) $\eta_r$–$\gamma$ curve with activity turned off for $\gamma \leq 1.4$ and turned on for $\gamma > 1.4$. The grey-shaded region indicates the high-viscosity jammed state. The light-green region indicates where activity is turned on. The black and red lines represent the $\eta_r$–$\gamma$ curves for systems without activity and with activity always on, respectively.
(c) Macroscopic friction coefficient $\mu$ as a function of the viscous number $J$ for different activity strengths.
(d) Suspension temperature $\Theta_s$ as a function of $J$.
(e) $\mu$ scaled by $\Theta_s$ shown as a function of $J$ for various activity strengths.}
\label{fig4}
\end{figure}

Data presented in Fig.~\ref{fig1} suggest that the most important parameter controlling the viscosity is the active force. We further explore, for fixed $C_a$ and $\tau_p$, how $F_a^m$ and $\eta_r^m$ depend on $\phi$. Fig.~\ref{fig2}(a) shows dependence of $\eta_r$ on $F_a$ for $C_a = 1$ and $\tau_p = 1$ over a range of $\phi$ values spanning above and below $\phi_J^{\mu_p = 1}$. Both $F_a^m$ and $\eta_r^m$ increase with $\phi$ and diverge close to $\phi_{\text{RCP}}$ as shown in Fig.~\ref{fig2}(b) where green and red solid lines represent the functional forms $F_a^m\left(\phi, \tau_p\right) = \mathcal{F}_1(\phi) \sim (\phi_J - \phi)^{-\alpha}$ and $\eta_r^m\left(\phi, \tau_p\right) = \mathcal{F}_2(\phi) \sim (\phi_J - \phi)^{-\beta}$, respectively, with $\alpha = 2.04$, $\beta = 2.64$, and $\phi_J = 0.636$. The Inset of (b) shows the collapse of data presented in (a) by scaling $F_a$ and $\eta_r$ using $\mathcal{F}_1(\phi)$ and $\mathcal{F}_2(\phi)$.  Similar dependence on $\tau_p$ is also expected but at $\phi < \phi_J^{\mu_p}$ since the effect of $\tau_p$ is weaker in this regime.  
\paragraph{Microscopic understanding of viscosity reduction.}
Next, we aim to understand the microscopic picture of how the presence of active particles leads to a significant reduction in viscosity. To do this, we study the dependence of three different microscopic quantities - mean number of frictional contacts $n_{fc} = n_{fc}^{total}/N$, force chain network, and the correlation length $\xi$ associated with velocity-velocity correlations - on different control parameters of active particles. Here, $n_{fc}^{total}$ is the total number of frictional contacts in the system. In Fig.~\ref{fig3}(a), the variation of frictional contacts with active force (bottom, in red) for $\phi = 0.56, C_a = 0.3$ and $\tau_p = 1$ and persistence time (top, in green) for $\phi = 0.56, C_a = 0.3$ and $F_a = 0.75$ is shown. The trend is similar to that observed in Fig.~\ref{fig1}(b) and (c), suggesting that the number of frictional contacts varies depending on the magnitude of $F_a$ and $\tau_p$, which in turn leads to changes in $\eta_r$. In the case of small $F_a$, the number of frictional contacts remains unchanged, indicating that the activity is not sufficient to break existing contacts or prevent their formation. However, as $F_a$ increases, contacts begin to break or fail to form. At large values, the number of contacts starts increasing again due to enhanced particle diffusion driven by strong active forcing. The dependence on $\tau_p$ is similar except for the plateau observed at small $F_a$ which is absent in this case. In Fig.~\ref{fig3}(b), the viscosity contributions from contact forces ($\eta_r^c$) and hydrodynamic forces ($\eta_r^h$) are shown. As frictional contacts break or fail to form, $\eta_r^c$ decreases, leading to an increase in purely hydrodynamic interactions and thus a higher $\eta_r^h$. We further explore the force chain network (Fig.~\ref{fig3}(e) and (f)) to illustrate the microscopic differences between passive and active systems at different $\phi$ but with the same $\eta_r$. Force chains are represented by lines connecting the centres of contacting particles, with the colour of each line indicating the magnitude of the contact force. For clarity, the same colour scale is used in all panels, with force magnitudes capped at a maximum of 0.5. When the force exceeds 0.5, it is truncated to this upper limit. We observe that the passive system with $\phi = 0.49$ exhibits long, branched, and anisotropic force chains. In contrast, the active system with $\phi = 0.56$, $C_a = 0.3$, $F_a = 0.75$ and $\tau_p = 1$ having the same $\eta_r$ as the passive system, displays shorter, more isotropic chains, although some contacts exhibit significantly higher force magnitudes compared to others. This suggests that in active systems, different regions are less correlated. To underpin this observation we study the correlation length scale $\xi$ associated with the velocity correlation function $c(r) = \frac{ \sum_{i \neq j} \vec{u}_i \cdot \vec{u}_j \, \delta\left(|r_{ij}| - r\right)}{\sum_{i \neq j} \delta\left(|r_{ij}| - r\right)}$~\cite{AnaelPRE2007, SinghPRL2020}, which decays exponentially as $c(r) \sim e^{-r/\xi}$. The decay of $\xi$ with increasing $F_a$, as shown in Fig.~\ref{fig3}(c), indicates that higher activity leads to less correlated systems. 
\paragraph{Tuning the jamming volume fraction.} 
The systematic reduction of contact numbers with increasing activity, by varying $C_a$, $F_a$ and $\tau_p$, suggests that it is possible to tune the shear jamming volume fraction with the help of active particles. Shear jamming occurs due to anisotropy in the microscopic structure, which leads to $n_{fc}> 4$. However, active particles introduce local straining in addition to global straining, which either breaks the force chains or prevents their formation and reduces structural anisotropy. In Fig.~\ref{fig4}(a), the data for the $\eta_r$ - $\phi$ curve for a passive system with $\mu_p = 1$ are shown by black circles. For this system, shear jamming occurs at $\phi_J^{\mu_p = 1} \sim 0.58$. The red points represent the active system with $F_a = 0.25$, $C_a = 0.3$, and $\tau_p = 1$, which introduces mild activity into the system. Such a system can reduce the viscosity at small $\phi$, but as $\phi$ increases, the effect of the active forcing becomes insignificant, and the $\eta_r - \gamma$ curve starts to approach that of the passive system, eventually converging with it. The viscosity then diverges at the same jamming volume fraction as the passive system. With a further increase of $F_a$ to 0.75 keeping $C_a$ and $\tau_p$ the same, the viscosity remains significantly lower than in the passive system at higher $\phi$, but begins to increase dramatically as $\phi$ approaches the jamming volume fraction. However, jamming occurs at $\phi > \phi_J^{\mu_p = 1}$. Increasing $F_a$ further from 0.75 to 3.0 shifts the jamming volume fraction close to 0.60. The data points for $F_a = 3$ and 90 with $C_a = 1$ suggest it is possible to push $\phi_J^{\mu_p = 1}$ to $\phi_{\text{RCP}}$. Note that with increasing $F_a$, $\eta_r$ at smaller $\phi$ is higher than what is obtained for smaller forces. This is consistent with the existence of a $\phi$-dependent $F_a^m$, as discussed in Fig.~\ref{fig1}. However for frictionless system $\mu_p = 0$ (Fig.~\ref{fig4} (b))the jamming volume fraction remains the same for active and passive systems in consistence with Ref.~\cite{NMC_ScAdv_2018,Pappu_ArXiv2025}. Moreover, the reduction of viscosity is negligible at smaller $\phi$  consistent with Ref.~\cite{NMC_ScAdv_2018} but in contrast with Ref.~\cite{Pappu_ArXiv2025}. 

We further explore the jamming-unjamming transition facilitated by activity for $\phi = 0.58$, $\mu_p = 1$. In the absence of activity, this system is jammed (in the hard-sphere limit) with a very large viscosity, $\eta_r \sim 4 \times 10^4$ (grey region), as shown by the black line in~\ref{fig4}(b). In the presence of activity with $F_a = 3$, $C_a = 1$, and $\tau_p = 1$, the system attains a low-viscosity steady state, shown by the red line. The green and blue lines are for $C_a = 1, F_a = 10$ and $\tau_p = 0.25$,  $C_a = 1, F_a = 10$ and $\tau_p = 0.5$ ,respectively. The activity is kept off in the regions $\gamma < 1.4$ where these systems get jammed but viscosity decreases when active is switched on (light green region). Interestingly we observe that, once the system is jammed, with $F_a = 3$, it can not be unjammed. It strongly suggests that if the activity is applied from the beginning of the deformation, a smaller $F_a$ can lead to fluidization. In contrast, once the system is jammed, a much larger value of $F_a$ is required to unjam it, suggesting the different roles of active particles in these two distinct scenarios. While for unjammed systems random motion of active particles prevents the formation of force chains, in the case of an already jammed system the active forcing reduces the viscosity by breaking the force chain which requires higher $F_a$.

\paragraph{Constitutive law.} Finally, we test the validity of the constitutive laws that describe the rheology of dense suspensions of granular particles, expressed in terms of three dimensionless numbers: the volume fraction $\phi$, the ratio of shear stress to normal stress $P$ known as macroscopic friction coefficient $\mu = \sigma_{xy}/P$, and the ratio of the viscous timescale to the shear timescale, known as viscous number $J = \eta_f \dot{\gamma}/P$. Such rheology is commonly known as the $\mu$–$J$ rheology. In Fig.~\ref{fig4}(d), the passive system (black circles) follows the constitutive law of the $\mu$–$J$ rheology, whereas the data for active systems do not. This is consistent with previous observations under inhomogeneous driving rates~\cite{NottBrady1994, migration3,NessGillissenPRL2020}, since locally, the active force induces local straining with a strain rate that differs from the global strain rate, thereby producing behaviour more akin to inhomogeneous flow. Recently, a novel dimensionless quantity called the suspension temperature, $\Theta_s = \eta_f \delta u/aP$, was introduced to provide a unified description of the rheology of both homogeneous and inhomogeneous suspensions, where $\delta u$ denotes the local velocity fluctuation~\cite{KimKamrin2020PRL,GranulartempVibration,BhowmikAndNessPRL,BhowmikAndNessPRL,BhowmikNessJoR2025}. $\Theta_s$ is computed by dividing the system into small blocks, computing locally, and then averaging over all blocks. Its variation as a function of $J$ is shown in (e) for systems with different activity strengths. In (f), $\mu$, appropriately scaled by $\Theta_s$, is plotted as a function of $J$. The quality of the data collapse suggests a new constitutive law, $\mu^{1.65} \Theta_s = \gamma \mathcal{F}(J)$, for the rheology of active dense suspensions where $\mathcal{F}(J)$ is the scaling function, and $\gamma = 1$ for active systems and 2 for passive systems.

\paragraph{Conclusion.}To conclude, we find that the presence of active particles significantly reduces the viscosity and shifts the jamming volume fraction to random close packing, $\phi_{\text{RCP}}$, for frictional systems. For frictionless systems, the reduction in viscosity is minimal and $\phi_J$ does not change, suggesting that tuning viscosity with active particles is more effective for frictional systems. Both the active force that leads to the minimum attained viscosity and the minimum viscosity itself diverge as $\phi$ approaches $\phi_J$. The microscopic picture of active systems is qualitatively different from that of passive systems. The active force required to unjam a shear-jammed system is significantly greater than the active force needed to maintain the system in a low-viscosity regime if the activity is sustained before deformation. Moreover, the rheology of active systems does not follow the conventional $\mu$–$J$ rheology. With the help of suspension temperature, a new constitutive law is defined for the rheology of active dense suspensions. Our findings will be useful for designing experiments to study the rheology of active suspensions. This study is limited to the rate-independent regime, and the effect of activity on phenomena such as shear thickening or shear thinning will be explored in future work. 
\begin{acknowledgments}
We are deeply grateful to Christopher Ness for encouraging us to pursue this research,  critical discussion, and support through funding. We would like to thank Smarajit Karmakar, Pappu Acharya and Anoop Mutneja for the various useful discussions.
\end{acknowledgments}

\bibliography{ALL}
\end{document}